\documentclass[12pt,preprint]{elsarticle}
\usepackage{amsmath,amssymb}
\usepackage{graphicx}
\graphicspath{{./}}
\usepackage{booktabs}
\usepackage{threeparttable}
\usepackage[table]{xcolor}
\usepackage{hyperref}
\usepackage{float}
\usepackage[margin=1in]{geometry}
\usepackage{setspace}
\usepackage{url}

\journal{Nuclear Science and Engineering}
\begin{document}
\begin{frontmatter}
\title{Rossi-$\alpha$ Benchmark Validation of a Static Alpha Eigenvalue Capability in OpenMC}
\author[llnl]{William Zywiec\corref{cor1}}
\ead{willzywiec@gmail.com}
\author[llnl,mit]{Alessandro Ingegno}
\author[llnl]{David Heinrichs}
\author[mit]{Benoit Forget}
\cortext[cor1]{Corresponding author.}
\address[llnl]{Lawrence Livermore National Laboratory, Livermore, CA 94550, USA}
\address[mit]{Massachusetts Institute of Technology, Cambridge, MA 02139, USA}

\begin{abstract}
A static alpha eigenvalue capability was implemented in a modified version of the open-source Monte Carlo radiation transport code OpenMC and validated against Rossi-$\alpha$ measurements from 21 delayed-critical benchmark experiments spanning fast, intermediate, and thermal systems with $^{233}$U, HEU, IEU, LEU, and plutonium fuels. An additional 33 subcritical configurations were analyzed to study the behavior of the alpha eigenvalue as a function of subcriticality. The effective delayed neutron fraction and the prompt neutron lifetime were both calculated using the iterated fission probability method, evaluated within the standard $k$-eigenvalue power iteration. Two forms of the alpha eigenvalue were calculated from these quantities using the point kinetics approximation: the delayed-critical alpha eigenvalue ($\alpha_{\mathrm{dc}} = -\beta_{\mathrm{eff}} / \ell_p$) and the static alpha eigenvalue ($\alpha_{\mathrm{static}} = (k_p - 1) / \ell_p$), which reflects the actual reactivity state of the model. The calculated $\alpha_{\mathrm{dc}}$ values are generally in good agreement with the measured Rossi-$\alpha$ values, with the closest agreement observed for the STACY thermal solution benchmarks. Subcritical extrapolation studies show that $\alpha_{\mathrm{dc}}$ remains approximately constant as the system is driven subcritical, while $\alpha_{\mathrm{static}}$ decreases with subcriticality and follows the available pulsed neutron source measurements. The results are broadly consistent with previously published values from MCNP5, Serpent~2, and OpenMC(TD); the differences that remain after accounting for the nuclear data library are attributable mainly to the methods used to calculate the underlying kinetics parameters.
\end{abstract}

\end{frontmatter}

\section{Introduction}\label{sec:intro}

The Rossi-$\alpha$ parameter characterizes the exponential rate of change in the prompt neutron population of a fissioning system near delayed critical~\cite{orndoff1957}. It is the prompt neutron decay constant or, equivalently, the fundamental-mode alpha eigenvalue of the time-dependent transport equation. Its dependence on both the effective delayed neutron fraction and the prompt neutron lifetime makes it particularly sensitive to the energy spectrum of the system and a useful validation observable for Monte Carlo radiation transport calculations.

In this work, two forms of the alpha eigenvalue are derived using the point kinetics approximation: the delayed-critical alpha eigenvalue ($\alpha_{\mathrm{dc}}$) and the static alpha eigenvalue ($\alpha_{\mathrm{static}}$). The delayed-critical alpha eigenvalue is the value at delayed critical regardless of the system's actual reactivity state and is the quantity measured in Rossi-$\alpha$ experiments~\cite{spriggs1999}. The static alpha eigenvalue reflects the actual reactivity state of the model and equals $\alpha_{\mathrm{dc}}$ only when $k_{\mathrm{eff}} = 1$.

Kiedrowski et al.~\cite{kiedrowski2011} implemented adjoint-weighted kinetics parameter tallies in MCNP5 using the iterated fission probability (IFP) method. Mosteller and Kiedrowski~\cite{mosteller2011} subsequently developed a 13-benchmark Rossi-$\alpha$ validation suite that has become a standard reference. Lepp\"anen et al.~\cite{leppanen2014} implemented a similar capability in Serpent~2 and validated it against 31 benchmarks. Romero-Barrientos et al.~\cite{romero2022} modified OpenMC to include time-dependent capabilities and calculated kinetics parameters for 18 benchmarks using the $k$-prompt method. Dorville et al.~\cite{dorville2025} have since implemented the IFP method in the released version of OpenMC, reporting adjoint-weighted $\beta_{\mathrm{eff}}$ and $\Lambda_{\mathrm{eff}}$ rather than the alpha eigenvalue.

This paper describes an independent implementation of a static alpha eigenvalue capability in a modified version of OpenMC~\cite{zywiec_openmc,romano2015}. The effective delayed neutron fraction ($\beta_{\mathrm{eff}}$) and the prompt neutron lifetime ($\ell_p$) are both calculated using the IFP method, and the alpha eigenvalue is calculated from these quantities using the point kinetics approximation. The results are compared with 21 delayed-critical Rossi-$\alpha$ values from the ICSBEP Handbook~\cite{icsbep} and other sources~\cite{csewg,tonoike2002,takano1985}. An additional 33 subcritical configurations~\cite{tonoike2002,takano1985} are analyzed to study the behavior of the alpha eigenvalue as a function of subcriticality.

\section{Theory and Implementation}\label{sec:theory}

The implementation evaluates $\beta_{\mathrm{eff}}$ and $\ell_p$ within the standard $k$-eigenvalue power iteration. The prompt neutron generation time ($\Lambda_p$) is also calculated for reporting purposes but is not used in the alpha eigenvalue calculation itself.

The point kinetics approximation assumes that the neutron flux is separable into spatial and temporal components:
\begin{equation}
\Psi(\mathbf{r}, \hat{\Omega}, E, t) = \psi(\mathbf{r}, \hat{\Omega}, E) \, T(t),
\end{equation}
where the temporal amplitude function takes the form $T(t) = T(0) \, e^{\alpha t}$. Two quantities are required to evaluate $\alpha$: $\beta_{\mathrm{eff}}$ and $\ell_p$.

Both are obtained from the IFP method~\cite{kiedrowski2011,nauchi2010}. Each fission neutron carries a rolling genealogy window of $N = 10$ generations, recording the lifetime of its ancestor $N$ generations earlier and whether that ancestor was born prompt or delayed.

The effective delayed neutron fraction ($\beta_{\mathrm{eff}}$) is the fraction of the asymptotic progeny weight that descends from delayed neutrons:
\begin{equation}
\beta_{\mathrm{eff}} = \frac{\displaystyle\sum_{i \in \mathcal{F}_d} w_i}{\displaystyle\sum_{i \in \mathcal{F}} w_i},
\label{eq:beff}
\end{equation}
where $\mathcal{F}$ is the set of fission events in the current generation caused by neutrons for which a full $N$-generation genealogy is available, $w_i$ is the weight of the neutron causing fission $i$, and $\mathcal{F}_d \subseteq \mathcal{F}$ restricts the numerator to those neutrons whose ancestor $N$ generations earlier was born delayed. The prompt multiplication factor ($k_p$) is then calculated as
\begin{equation}
k_p = k_{\mathrm{eff}} (1 - \beta_{\mathrm{eff}}),
\label{eq:kp}
\end{equation}
where $k_{\mathrm{eff}}$ is the effective multiplication factor.

The prompt neutron lifetime ($\ell_p$) is the mean time between the birth of a prompt neutron and the fission it causes. It is evaluated from the same genealogy:
\begin{equation}
\ell_p = \frac{\displaystyle\sum_{i \in \mathcal{F}_p} \ell_i^{(N)} \cdot w_i}{\displaystyle\sum_{i \in \mathcal{F}_p} w_i},
\label{eq:ellp_ifp}
\end{equation}
where $\ell_i^{(N)}$ is the lifetime of neutron $i$'s ancestor emitted $N$ generations earlier and $\mathcal{F}_p \subseteq \mathcal{F}$ restricts both sums to those neutrons whose ancestor was born prompt.

Both quantities are therefore adjoint-weighted~\cite{bell1970}: each fission neutron contributes in proportion to its importance to the asymptotic fission chain rather than simply being counted. The progeny weight $N$ generations later serves as the importance estimate, so no explicit solution of the adjoint transport equation is required.

When a system is delayed-critical ($k_{\mathrm{eff}} = 1$, $k_p = 1 - \beta_{\mathrm{eff}}$), the prompt neutron population decays because $k_p < 1$. Given $\beta_{\mathrm{eff}}$ and $\ell_p$, the delayed-critical alpha eigenvalue ($\alpha_{\mathrm{dc}}$) is calculated as
\begin{equation}
\alpha_{\mathrm{dc}} = \frac{-\beta_{\mathrm{eff}}}{\ell_p}.
\label{eq:alpha_dc}
\end{equation}
For a model where the calculated $k_{\mathrm{eff}}$ departs from unity, replacing $-\beta_{\mathrm{eff}}$ with $k_p - 1$ gives the static alpha eigenvalue ($\alpha_{\mathrm{static}}$):
\begin{equation}
\alpha_{\mathrm{static}} = \frac{k_p - 1}{\ell_p}.
\label{eq:alpha_static}
\end{equation}
The uncertainty on $\alpha_{\mathrm{dc}}$ depends on $\sigma(\beta_{\mathrm{eff}})$ and $\sigma(\ell_p)$. The uncertainty on $\alpha_{\mathrm{static}}$ additionally depends on $\sigma(k_{\mathrm{eff}})$ through $k_p$.

The prompt neutron generation time ($\Lambda_p$) is related to the prompt lifetime by
\begin{equation}
\Lambda_p = \frac{\ell_p}{k_p}.
\label{eq:gentime}
\end{equation}
Its uncertainty is propagated from $\sigma(\ell_p)$ and $\sigma(k_p)$.

The alpha eigenvalues calculated using Eqs.~(\ref{eq:alpha_dc}) and~(\ref{eq:alpha_static}) are derived from $k$-eigenvalue tallies rather than directly simulated time-dependent eigenvalues. For systems near delayed critical, the point kinetics approximation is well justified and $\alpha_{\mathrm{dc}}$ closely approximates the true time-dependent alpha eigenvalue. The linear relationship between $\alpha_{\mathrm{static}}$ and reactivity implied by Eq.~(\ref{eq:alpha_static}) holds only in the vicinity of delayed critical; as systems become more deeply subcritical (below $k_{\mathrm{eff}} \approx 0.95$), higher spatial and energy modes contribute more to the time-dependent response~\cite{spriggs1997,mckenzie2018}, and both measured and calculated alpha eigenvalues progressively depart from the fundamental-mode point kinetics behavior. The same caveat applies above delayed critical, where $\beta_{\mathrm{eff}}$ and $\ell_p$ themselves vary with configuration so that $\alpha_{\mathrm{static}}$ is not strictly linear in reactivity~\cite{orndoff1957}. The configurations analyzed in this work range from $k_{\mathrm{eff}} \approx 0.90$ to 1.009. Quantifying the limits of the present approach at deeper subcriticality is left for future work.

Enabling the static alpha eigenvalue capability increases the active-batch simulation time relative to a standard $k$-eigenvalue calculation by a factor of 3.5 for a fast benchmark (Godiva, from 15.3 to 54.3~s) and a factor of 1.9 for a thermal solution benchmark (STACY-029, from 276.1 to 511.9~s). All four calculations were run with identical settings on the same 22-core workstation using 22 threads. The penalty is smaller for the thermal system because the IFP bookkeeping is a fixed cost per generation, while the transport work per history is far greater in a thermal solution than in a bare metal sphere.

\section{Benchmarks}\label{sec:benchmarks}

The 21 delayed-critical benchmarks are listed in Table~\ref{tab:benchmarks}. The suite spans $^{233}$U, HEU, IEU, LEU, and plutonium fuels, with fast, intermediate, and thermal spectra. Reflector materials include natural uranium, depleted uranium, thorium, copper, graphite, and water. All OpenMC calculations were performed using ENDF/B-VIII.0 cross sections with 50 inactive and 4,950 active batches of 50,000 source neutrons.

\begin{table}[htbp]
\centering
\caption{Delayed-critical benchmarks.}
\label{tab:benchmarks}
\footnotesize
\begin{tabular}{@{}lllll@{}}
\toprule
Name & ICSBEP ID & Fuel & Spectrum & Description \\
\midrule
Godiva       & HMF001   & HEU       & Fast         & Bare sphere of HEU \\
Flattop-25   & HMF028   & HEU       & Fast         & HEU sphere, nat-U reflector \\
Zeus-1       & HMI006-1 & HEU       & Intermediate & HEU/graphite, Cu reflector \\
Zeus-5       & HMF073-1 & HEU       & Fast         & HEU plates, Cu reflector \\
Zeus-6       & HMF072-1 & HEU       & Fast         & HEU/steel, Cu reflector \\
Big Ten      & IMF007-1 & IEU       & Fast         & IEU cylinders, DU reflector \\
Winco        & HST038-1 & HEU soln. & Thermal      & Interacting slab tanks, UNH soln. \\
Jezebel-Pu   & PMF001   & Pu        & Fast         & Bare sphere of Pu \\
Flattop-Pu   & PMF006   & Pu        & Fast         & Pu sphere, nat-U reflector \\
Thor         & PMF008   & Pu        & Fast         & Pu sphere, Th reflector \\
Jezebel-233  & UMF001-1 & $^{233}$U & Fast         & Bare sphere of $^{233}$U \\
Flattop-23   & UMF006-1 & $^{233}$U & Fast         & $^{233}$U sphere, nat-U reflector \\
RA-6         & ICT014-2 & IEU       & Thermal      & Open pool research reactor \\
SHE-8        & ---      & LEU       & Thermal      & Graphite-matrix LEU fuel \\
SHEBA-II     & LST001-1 & LEU soln. & Thermal      & Cyl.\ tank, UO$_2$F$_2$ soln. \\
STACY-029    & LST004-2 & LEU soln. & Thermal      & Water-refl.\ cyl.\ tank, UNH soln. \\
STACY-033    & LST004-3 & LEU soln. & Thermal      & Water-refl.\ cyl.\ tank, UNH soln. \\
STACY-046    & LST004-5 & LEU soln. & Thermal      & Water-refl.\ cyl.\ tank, UNH soln. \\
STACY-030    & LST007-2 & LEU soln. & Thermal      & Unreflected cyl.\ tank, UNH soln. \\
STACY-125    & LST016-3 & LEU soln. & Thermal      & Water-refl.\ slab tanks, UNH soln. \\
STACY-215    & LST021-1 & LEU soln. & Thermal      & Unreflected cyl.\ tank, UNH soln. \\
\bottomrule
\end{tabular}
\end{table}

An additional 33 subcritical configurations from the SHE-8 and STACY benchmarks were generated by absorber insertion (SHE-8) or solution height reduction (STACY). These are summarized in Table~\ref{tab:subcrit_benchmarks}.

\begin{table}[htbp]
\centering
\caption{Subcritical configurations.}
\label{tab:subcrit_benchmarks}
\footnotesize
\begin{tabular}{@{}lccl@{}}
\toprule
Parent & Cases & $k_{\mathrm{eff}}$ range & Method \\
\midrule
SHE-8    & 5  & 0.900--0.973 & Gray/black absorber insertion \\
STACY-029 & 5 & 0.977--1.000 & Solution height reduction \\
STACY-033 & 3 & 0.990--0.998 & Solution height reduction \\
STACY-046 & 5 & 0.966--1.001 & Solution height reduction \\
STACY-030 & 5 & 0.969--0.996 & Solution height reduction \\
STACY-125 & 5 & 0.990--1.004 & Solution height reduction \\
STACY-215 & 5 & 0.984--0.995 & Solution height reduction \\
\bottomrule
\end{tabular}
\end{table}

Calculated kinetics parameters for the 21 delayed-critical benchmarks are given in Table~\ref{tab:kinetics}; those for the 33 subcritical configurations are presented alongside the measured values in Tables~\ref{tab:she8_subcrit} through~\ref{tab:stacy215}.

\begin{table}[htbp]
\centering
\caption{Calculated kinetics parameters at delayed critical.}
\label{tab:kinetics}
\footnotesize
\begin{tabular}{@{}lcccccc@{}}
\toprule
Name & OpenMC $k_{\mathrm{eff}}^*$ & OpenMC $k_p^*$ & OpenMC $\beta_{\mathrm{eff}}$ (pcm) & OpenMC $\ell_p$ & OpenMC $\Lambda_p$ & Units \\
\midrule
Godiva      & 1.00013 & 0.99363 & $650 \pm 2$ & $5.594 \pm 0.003$ & $5.630 \pm 0.003$ & $\times 10^{-9}$ s \\
Flattop-25  & 1.00083 & 0.99392 & $690 \pm 2$ & $1.713 \pm 0.001$ & $1.723 \pm 0.001$ & $\times 10^{-8}$ s \\
Zeus-1      & 0.99576 & 0.98839 & $740 \pm 2$ & $2.000 \pm 0.002$ & $2.024 \pm 0.002$ & $\times 10^{-6}$ s \\
Zeus-5      & 1.00204 & 0.99526 & $677 \pm 2$ & $6.438 \pm 0.007$ & $6.469 \pm 0.007$ & $\times 10^{-8}$ s \\
Zeus-6      & 1.00420 & 0.99741 & $676 \pm 2$ & $1.671 \pm 0.001$ & $1.675 \pm 0.001$ & $\times 10^{-7}$ s \\
Big Ten     & 1.00429 & 0.99705 & $721 \pm 2$ & $6.055 \pm 0.003$ & $6.073 \pm 0.003$ & $\times 10^{-8}$ s \\
Winco       & 0.99465 & 0.98645 & $824 \pm 2$ & $7.359 \pm 0.003$ & $7.460 \pm 0.004$ & $\times 10^{-6}$ s \\
Jezebel-Pu  & 0.99981 & 0.99794 & $187 \pm 1$ & $2.818 \pm 0.002$ & $2.824 \pm 0.002$ & $\times 10^{-9}$ s \\
Flattop-Pu  & 0.99963 & 0.99683 & $280 \pm 2$ & $1.303 \pm 0.001$ & $1.307 \pm 0.001$ & $\times 10^{-8}$ s \\
Thor        & 0.99741 & 0.99533 & $209 \pm 1$ & $9.860 \pm 0.010$ & $9.906 \pm 0.010$ & $\times 10^{-9}$ s \\
Jezebel-233 & 1.00052 & 0.99755 & $297 \pm 1$ & $2.723 \pm 0.001$ & $2.730 \pm 0.001$ & $\times 10^{-9}$ s \\
Flattop-23  & 0.99999 & 0.99622 & $377 \pm 2$ & $1.247 \pm 0.001$ & $1.252 \pm 0.001$ & $\times 10^{-8}$ s \\
RA-6        & 1.00477 & 0.99725 & $749 \pm 2$ & $4.308 \pm 0.003$ & $4.320 \pm 0.003$ & $\times 10^{-5}$ s \\
SHE-8       & 1.00377 & 0.99670 & $704 \pm 2$ & $1.083 \pm 0.001$ & $1.087 \pm 0.001$ & $\times 10^{-3}$ s \\
SHEBA-II    & 1.00899 & 1.00134 & $757 \pm 2$ & $3.992 \pm 0.002$ & $3.987 \pm 0.002$ & $\times 10^{-5}$ s \\
STACY-029   & 1.00190 & 0.99459 & $730 \pm 2$ & $5.927 \pm 0.002$ & $5.959 \pm 0.002$ & $\times 10^{-5}$ s \\
STACY-033   & 0.99985 & 0.99260 & $726 \pm 2$ & $6.215 \pm 0.002$ & $6.261 \pm 0.003$ & $\times 10^{-5}$ s \\
STACY-046   & 1.00210 & 0.99496 & $713 \pm 2$ & $6.690 \pm 0.003$ & $6.724 \pm 0.003$ & $\times 10^{-5}$ s \\
STACY-030   & 0.99758 & 0.99023 & $736 \pm 2$ & $5.799 \pm 0.002$ & $5.856 \pm 0.002$ & $\times 10^{-5}$ s \\
STACY-125   & 1.00482 & 0.99735 & $744 \pm 2$ & $4.866 \pm 0.002$ & $4.879 \pm 0.002$ & $\times 10^{-5}$ s \\
STACY-215   & 0.99784 & 0.99065 & $721 \pm 2$ & $6.622 \pm 0.003$ & $6.685 \pm 0.003$ & $\times 10^{-5}$ s \\
\bottomrule
\multicolumn{7}{@{}l}{\footnotesize $^*$Statistical uncertainty on $k_{\mathrm{eff}}$ and $k_p$ is $< 0.0001$ for all cases.} \\
\end{tabular}
\end{table}

\section{Results}\label{sec:results}

\subsection{Alpha Eigenvalue Comparison}

\begin{figure}[htbp]
\centering
\includegraphics[width=0.85\textwidth]{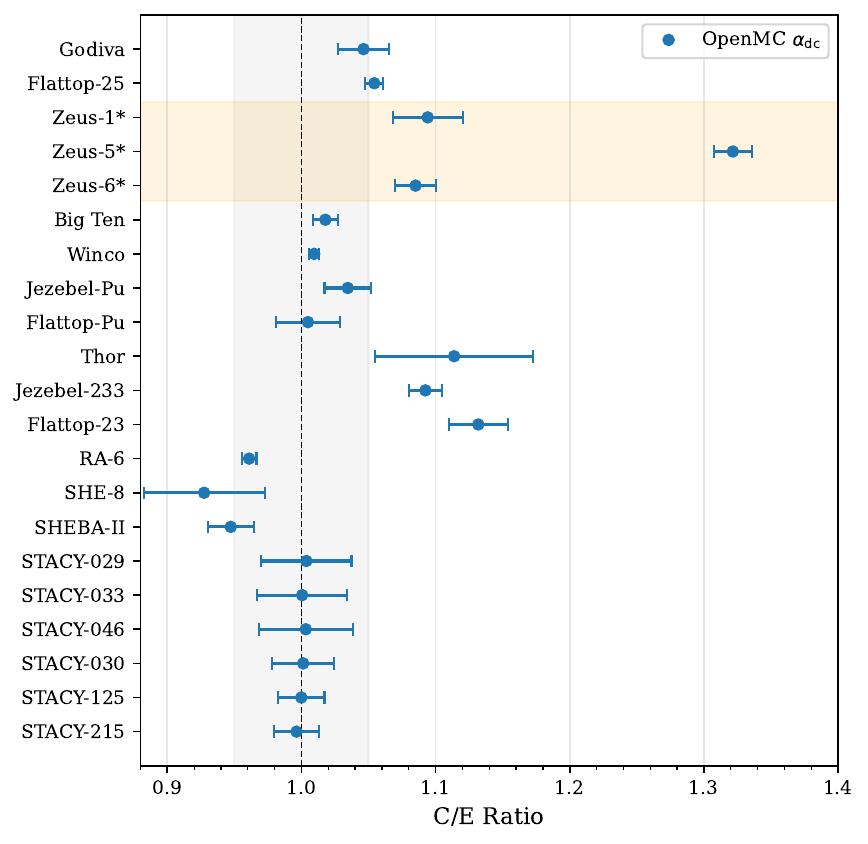}
\caption{C/E for $\alpha_{\mathrm{dc}}$ across all 21 benchmarks ($^\ast$See Section~\ref{sec:zeus}). The gray vertical band spans C/E $= 1 \pm 0.05$ ($\pm 5\%$) and is provided for visual reference only; the orange horizontal band marks the three Zeus benchmarks. Error bars represent the combined statistical uncertainty of the OpenMC calculation and the reported measurement uncertainty, propagated in quadrature.}
\label{fig:ce_ratio}
\end{figure}

The comparison of measured and calculated $\alpha_{\mathrm{dc}}$ values is presented in Table~\ref{tab:alpha_comparison} and Fig.~\ref{fig:ce_ratio}. For the six STACY thermal solution benchmarks, calculated-to-experimental (C/E) ratios range from 0.996 to 1.004. Among the other thermal systems, Winco agrees to within 1.0\%, while RA-6, SHEBA-II, and SHE-8 are underpredicted by 3.9\%, 5.3\%, and 7.3\%, respectively. The fast metal systems generally agree within 5--10\%, comparable to results reported by Mosteller and Kiedrowski~\cite{mosteller2011} and Lepp\"anen et al.~\cite{leppanen2014}, but are consistently overpredicted, most strongly for the fertile-reflected Flattop-23 (13\%) and Thor (11\%). The three Zeus benchmarks remain outliers, as discussed in Section~\ref{sec:zeus}.

For benchmarks where the calculated $k_{\mathrm{eff}}$ is close to unity, $\alpha_{\mathrm{dc}}$ and $\alpha_{\mathrm{static}}$ are nearly identical; when $k_{\mathrm{eff}}$ departs from unity, the two diverge. SHEBA-II is the extreme case: the model is prompt-supercritical ($k_p = 1.00134$, Table~\ref{tab:kinetics}), so the prompt neutron population grows rather than decays and $\alpha_{\mathrm{static}}$ is positive.

\begin{table}[htbp]
\centering
\footnotesize
\begin{threeparttable}
\caption{Measured and calculated $\alpha_{\mathrm{dc}}$ at delayed critical.$^*$}
\label{tab:alpha_comparison}
\begin{tabular}{@{}lccrcc@{}}
\toprule
Name & Measured $\alpha_{\mathrm{dc}}$~\cite{icsbep,csewg,tonoike2002,takano1985} & OpenMC $\alpha_{\mathrm{dc}}$ & C/E & OpenMC $\alpha_{\mathrm{static}}$ & Units (s$^{-1}$) \\
\midrule
Godiva       & $-1.110 \pm 0.020$ & $-1.1614 \pm 0.0043$ & 1.046 & $-1.1391 \pm 0.0089$ & $\times 10^{6}$ \\
Flattop-25   & $-0.382 \pm 0.002$ & $-0.4028 \pm 0.0015$ & 1.054 & $-0.3549 \pm 0.0032$ & $\times 10^{6}$ \\
Zeus-1       & $-0.338 \pm 0.008$ & $-0.3698 \pm 0.0013$ & 1.094 & $-0.5804 \pm 0.0031$ & $\times 10^{4}$ \\
Zeus-5       & $-7.960 \pm 0.080$ & $-10.5200 \pm 0.0384$ & 1.322 & $-7.3696 \pm 0.0858$ & $\times 10^{4}$ \\
Zeus-6       & $-3.730 \pm 0.050$ & $-4.0474 \pm 0.0143$ & 1.085 & $-1.5485 \pm 0.0324$ & $\times 10^{4}$ \\
Big Ten      & $-0.117 \pm 0.001$ & $-0.1191 \pm 0.0004$ & 1.018 & $-0.0487 \pm 0.0008$ & $\times 10^{6}$ \\
Winco        & $-11.090 \pm 0.030$ & $-11.1955 \pm 0.0305$ & 1.010 & $-18.4123 \pm 0.0918$ & $\times 10^{2}$ \\
Jezebel-Pu   & $-0.640 \pm 0.010$ & $-0.6621 \pm 0.0040$ & 1.035 & $-0.7296 \pm 0.0151$ & $\times 10^{6}$ \\
Flattop-Pu   & $-0.214 \pm 0.005$ & $-0.2150 \pm 0.0012$ & 1.005 & $-0.2433 \pm 0.0040$ & $\times 10^{6}$ \\
Thor         & $-0.190 \pm 0.010$ & $-0.2116 \pm 0.0013$ & 1.114 & $-0.4734 \pm 0.0047$ & $\times 10^{6}$ \\
Jezebel-233  & $-1.000 \pm 0.010$ & $-1.0924 \pm 0.0055$ & 1.092 & $-0.9006 \pm 0.0158$ & $\times 10^{6}$ \\
Flattop-23   & $-0.267 \pm 0.005$ & $-0.3022 \pm 0.0014$ & 1.132 & $-0.3033 \pm 0.0042$ & $\times 10^{6}$ \\
RA-6         & $-1.808 \pm 0.009$ & $-1.7375 \pm 0.0049$ & 0.961 & $-0.6382 \pm 0.0152$ & $\times 10^{2}$ \\
SHE-8        & $-7.010 \pm 0.340$ & $-6.5014 \pm 0.0186$ & 0.927 & $-3.0455 \pm 0.0612$ & $\times 10^{0}$ \\
SHEBA-II     & $-2.003 \pm 0.036$ & $-1.8974 \pm 0.0053$ & 0.947 & $\phantom{-}0.3364 \pm 0.0163$ & $\times 10^{2}$ \\
STACY-029    & $-1.227 \pm 0.041$ & $-1.2315 \pm 0.0035$ & 1.004 & $-0.9131 \pm 0.0104$ & $\times 10^{2}$ \\
STACY-033    & $-1.167 \pm 0.039$ & $-1.1676 \pm 0.0034$ & 1.001 & $-1.1910 \pm 0.0098$ & $\times 10^{2}$ \\
STACY-046    & $-1.062 \pm 0.037$ & $-1.0654 \pm 0.0031$ & 1.003 & $-0.7532 \pm 0.0089$ & $\times 10^{2}$ \\
STACY-030    & $-1.268 \pm 0.029$ & $-1.2696 \pm 0.0036$ & 1.001 & $-1.6846 \pm 0.0106$ & $\times 10^{2}$ \\
STACY-125    & $-1.528 \pm 0.026$ & $-1.5279 \pm 0.0044$ & 1.000 & $-0.5453 \pm 0.0130$ & $\times 10^{2}$ \\
STACY-215    & $-1.092 \pm 0.018$ & $-1.0880 \pm 0.0032$ & 0.996 & $-1.4125 \pm 0.0092$ & $\times 10^{2}$ \\
\bottomrule
\end{tabular}
\begin{tablenotes}[flushleft]
\footnotesize
\item[]$^*$For models where the calculated $k_{\mathrm{eff}}$ departs from unity (Table~\ref{tab:kinetics}), the comparison is affected by the reactivity bias of the model (see Section~\ref{sec:zeus}).
\end{tablenotes}
\end{threeparttable}
\end{table}

\subsection{Comparison with Published Results}

Tables~\ref{tab:literature}--\ref{tab:lit_lambda} compare the OpenMC results from this work with previously published values from MCNP5~\cite{mosteller2011}, Serpent~2~\cite{leppanen2014}, and OpenMC(TD)~\cite{romero2022}. The four sets of results were obtained with different methods for calculating the underlying kinetics parameters and with different nuclear data libraries: ENDF/B-VIII.0 in this work, ENDF/B-VII.0 for Mosteller and Kiedrowski, JEFF-3.1.1 for Lepp\"anen et al., and ENDF/B-VII.1 for Romero-Barrientos et al. Mosteller and Kiedrowski report only $\alpha_{\mathrm{dc}}$ and do not tabulate $\beta_{\mathrm{eff}}$ or $\Lambda_p$ separately.

\begin{table}[htbp]
\centering
\caption{$\alpha_{\mathrm{dc}}$ comparison with published values.}
\label{tab:literature}
\footnotesize
\begin{tabular}{@{}lccccc@{}}
\toprule
Name & OpenMC $\alpha_{\mathrm{dc}}$ & MCNP5 $\alpha_{\mathrm{dc}}$~\cite{mosteller2011} & Serpent~2 $\alpha_{\mathrm{dc}}$~\cite{leppanen2014} & OpenMC(TD) $\alpha_{\mathrm{dc}}$~\cite{romero2022} & Units (s$^{-1}$) \\
\midrule
Godiva       & $-1.1614 \pm 0.0043$ & $-1.130 \pm 0.010$ & --- & $-1.110 \pm 0.020$ & $\times 10^{6}$ \\
Flattop-25   & $-0.4028 \pm 0.0015$ & $-0.397 \pm 0.002$ & --- & --- & $\times 10^{6}$ \\
Zeus-1       & $-0.3698 \pm 0.0013$ & $-0.363 \pm 0.002$ & --- & --- & $\times 10^{4}$ \\
Zeus-5       & $-10.5200 \pm 0.0384$ & $-10.760 \pm 0.070$ & --- & --- & $\times 10^{4}$ \\
Zeus-6       & $-4.0474 \pm 0.0143$ & $-4.140 \pm 0.030$ & --- & --- & $\times 10^{4}$ \\
Big Ten      & $-0.1191 \pm 0.0004$ & $-0.118 \pm 0.001$ & $-0.120 \pm 0.001$ & $-0.110 \pm 0.002$ & $\times 10^{6}$ \\
Winco        & $-11.1955 \pm 0.0305$ & --- & $-11.220 \pm 0.050$ & --- & $\times 10^{2}$ \\
Jezebel-Pu   & $-0.6621 \pm 0.0040$ & $-0.650 \pm 0.010$ & --- & $-0.628 \pm 0.028$ & $\times 10^{6}$ \\
Flattop-Pu   & $-0.2150 \pm 0.0012$ & $-0.210 \pm 0.003$ & --- & --- & $\times 10^{6}$ \\
Thor         & $-0.2116 \pm 0.0013$ & $-0.200 \pm 0.010$ & --- & --- & $\times 10^{6}$ \\
Jezebel-233  & $-1.0924 \pm 0.0055$ & $-1.080 \pm 0.010$ & --- & --- & $\times 10^{6}$ \\
Flattop-23   & $-0.3022 \pm 0.0014$ & $-0.302 \pm 0.004$ & --- & $-0.268 \pm 0.016$ & $\times 10^{6}$ \\
RA-6         & $-1.7375 \pm 0.0049$ & --- & --- & $-1.724 \pm 0.027$ & $\times 10^{2}$ \\
SHE-8        & $-6.5014 \pm 0.0186$ & --- & $-6.428 \pm 0.028$ & --- & $\times 10^{0}$ \\
SHEBA-II     & $-1.8974 \pm 0.0053$ & --- & $-2.113 \pm 0.009$ & --- & $\times 10^{2}$ \\
STACY-029    & $-1.2315 \pm 0.0035$ & --- & $-1.263 \pm 0.006$ & $-1.154 \pm 0.014$ & $\times 10^{2}$ \\
STACY-033    & $-1.1676 \pm 0.0034$ & --- & $-1.186 \pm 0.005$ & $-1.101 \pm 0.012$ & $\times 10^{2}$ \\
STACY-046    & $-1.0654 \pm 0.0031$ & --- & $-1.091 \pm 0.005$ & $-1.005 \pm 0.015$ & $\times 10^{2}$ \\
STACY-030    & $-1.2696 \pm 0.0036$ & --- & $-1.285 \pm 0.005$ & $-1.197 \pm 0.014$ & $\times 10^{2}$ \\
STACY-125    & $-1.5279 \pm 0.0044$ & --- & $-1.566 \pm 0.007$ & $-1.464 \pm 0.017$ & $\times 10^{2}$ \\
STACY-215    & $-1.0880 \pm 0.0032$ & --- & $-1.109 \pm 0.005$ & --- & $\times 10^{2}$ \\
\bottomrule
\end{tabular}
\end{table}

\begin{table}[htbp]
\centering
\caption{$\beta_{\mathrm{eff}}$ comparison (pcm).}
\label{tab:lit_beff}
\small
\begin{tabular}{@{}lccc@{}}
\toprule
Name & OpenMC $\beta_{\mathrm{eff}}$ & Serpent~2 $\beta_{\mathrm{eff}}$~\cite{leppanen2014} & OpenMC(TD) $\beta_{\mathrm{eff}}$~\cite{romero2022} \\
\midrule
Godiva       & $650 \pm 2$ & $644 \pm 3$ & $633 \pm 7$ \\
Flattop-25   & $690 \pm 2$ & $693 \pm 3$ & $688 \pm 8$ \\
Zeus-1       & $740 \pm 2$ & --- & --- \\
Zeus-5       & $677 \pm 2$ & --- & --- \\
Zeus-6       & $676 \pm 2$ & --- & --- \\
Big Ten      & $721 \pm 2$ & $731 \pm 3$ & $696 \pm 7$ \\
Winco        & $824 \pm 2$ & $847 \pm 4$ & --- \\
Jezebel-Pu   & $187 \pm 1$ & $189 \pm 2$ & $181 \pm 7$ \\
Flattop-Pu   & $280 \pm 2$ & $285 \pm 2$ & $280 \pm 8$ \\
Thor         & $209 \pm 1$ & --- & --- \\
Jezebel-233  & $297 \pm 1$ & $290 \pm 2$ & $293 \pm 7$ \\
Flattop-23   & $377 \pm 2$ & $378 \pm 3$ & $357 \pm 8$ \\
RA-6         & $749 \pm 2$ & --- & $769 \pm 14$ \\
SHE-8        & $704 \pm 2$ & $719 \pm 3$ & --- \\
SHEBA-II     & $757 \pm 2$ & $791 \pm 3$ & --- \\
STACY-029    & $730 \pm 2$ & $752 \pm 3$ & $713 \pm 8$ \\
STACY-033    & $726 \pm 2$ & $741 \pm 3$ & $718 \pm 8$ \\
STACY-046    & $713 \pm 2$ & $733 \pm 3$ & $709 \pm 8$ \\
STACY-030    & $736 \pm 2$ & $752 \pm 3$ & $736 \pm 8$ \\
STACY-125    & $744 \pm 2$ & $764 \pm 3$ & $753 \pm 8$ \\
STACY-215    & $721 \pm 2$ & $740 \pm 3$ & --- \\
\bottomrule
\end{tabular}
\end{table}

\begin{table}[htbp]
\centering
\caption{$\Lambda_p$ comparison.}
\label{tab:lit_lambda}
\footnotesize
\begin{tabular}{@{}lcccl@{}}
\toprule
Name & OpenMC $\Lambda_p$ & Serpent~2 $\Lambda_p$~\cite{leppanen2014} & OpenMC(TD) $\Lambda_p$~\cite{romero2022} & Units \\
\midrule
Godiva       & $5.630 \pm 0.003$ & $5.711 \pm 0.004$ & $5.699 \pm 0.095$ & $\times 10^{-9}$ s \\
Flattop-25   & $1.723 \pm 0.001$ & $1.743 \pm 0.002$ & $1.936 \pm 0.006$ & $\times 10^{-8}$ s \\
Zeus-1       & $2.024 \pm 0.002$ & --- & --- & $\times 10^{-6}$ s \\
Zeus-5       & $6.469 \pm 0.007$ & --- & --- & $\times 10^{-8}$ s \\
Zeus-6       & $1.675 \pm 0.001$ & --- & --- & $\times 10^{-7}$ s \\
Big Ten      & $6.073 \pm 0.003$ & $6.147 \pm 0.004$ & $6.338 \pm 0.010$ & $\times 10^{-8}$ s \\
Winco        & $7.460 \pm 0.004$ & $7.553 \pm 0.004$ & --- & $\times 10^{-6}$ s \\
Jezebel-Pu   & $2.824 \pm 0.002$ & $2.806 \pm 0.002$ & $2.879 \pm 0.060$ & $\times 10^{-9}$ s \\
Flattop-Pu   & $1.307 \pm 0.001$ & $1.295 \pm 0.002$ & $1.394 \pm 0.008$ & $\times 10^{-8}$ s \\
Thor         & $9.906 \pm 0.010$ & --- & --- & $\times 10^{-9}$ s \\
Jezebel-233  & $2.730 \pm 0.001$ & $2.652 \pm 0.002$ & $2.760 \pm 0.018$ & $\times 10^{-9}$ s \\
Flattop-23   & $1.252 \pm 0.001$ & $1.243 \pm 0.002$ & $1.330 \pm 0.007$ & $\times 10^{-8}$ s \\
RA-6         & $4.320 \pm 0.003$ & --- & $3.757 \pm 0.027$ & $\times 10^{-5}$ s \\
SHE-8        & $1.087 \pm 0.001$ & $1.121 \pm 0.001$ & --- & $\times 10^{-3}$ s \\
SHEBA-II     & $3.987 \pm 0.002$ & $3.747 \pm 0.001$ & --- & $\times 10^{-5}$ s \\
STACY-029    & $5.959 \pm 0.002$ & $5.956 \pm 0.002$ & $6.147 \pm 0.038$ & $\times 10^{-5}$ s \\
STACY-033    & $6.261 \pm 0.003$ & $6.271 \pm 0.003$ & $6.456 \pm 0.038$ & $\times 10^{-5}$ s \\
STACY-046    & $6.724 \pm 0.003$ & $6.718 \pm 0.003$ & $6.993 \pm 0.039$ & $\times 10^{-5}$ s \\
STACY-030    & $5.856 \pm 0.002$ & $5.857 \pm 0.002$ & $5.860 \pm 0.032$ & $\times 10^{-5}$ s \\
STACY-125    & $4.879 \pm 0.002$ & $4.880 \pm 0.002$ & $5.111 \pm 0.033$ & $\times 10^{-5}$ s \\
STACY-215    & $6.685 \pm 0.003$ & $6.680 \pm 0.003$ & --- & $\times 10^{-5}$ s \\
\bottomrule
\end{tabular}
\end{table}

To quantify the effect of using different nuclear data libraries, the complete benchmark suite was recalculated with ENDF/B-VII.0 and ENDF/B-VII.1 cross sections using the same code, models, and run settings. The results of this intercomparison are presented in Tables~\ref{tab:library_mcnp} and~\ref{tab:library_td}. Matching the nuclear data library improves agreement with both sets of published $\alpha_{\mathrm{dc}}$ values. Relative to the MCNP5 values of Mosteller and Kiedrowski~\cite{mosteller2011}, the mean absolute difference in $\alpha_{\mathrm{dc}}$ falls from 2.1\% to 1.2\%; relative to the OpenMC(TD) values of Romero-Barrientos et al.~\cite{romero2022}, it falls from 6.1\% to 4.7\%.

Mosteller and Kiedrowski~\cite{mosteller2011} calculate the underlying kinetics parameters with the same adjoint-weighted IFP method used in this work, and their values agree most closely with the present results. The larger differences with OpenMC(TD) reflect the different methods used to obtain these parameters. Romero-Barrientos et al.~\cite{romero2022} calculate $\beta_{\mathrm{eff}}$ using the $k$-prompt method ($\beta_{\mathrm{eff}} = 1 - k_p/k_{\mathrm{eff}}$), which is not adjoint-weighted, and obtain $\Lambda_p$ by fitting an exponential to the decay of the neutron population in a fixed-source calculation on a slightly subcritical configuration. For the benchmarks with published OpenMC(TD) values (Tables~\ref{tab:lit_beff} and~\ref{tab:lit_lambda}), their $\beta_{\mathrm{eff}}$ values differ from the adjoint-weighted values of this work by 1.9\% on average while their $\Lambda_p$ values differ by 4.7\%. A decay constant fitted to a driven subcritical system is sensitive to the source distribution and to the fitting interval, since higher spatial and energy modes contribute to the early-time response; Romero-Barrientos et al.~\cite{romero2022} report differences of up to 8.1\% between their fitted $\Lambda_p$ and the adjoint-weighted MCNP values for the same benchmarks. The fitted $\Lambda_p$, rather than $\beta_{\mathrm{eff}}$, is therefore the dominant contributor to the remaining differences with OpenMC(TD).

\begin{table}[htbp]
\centering
\caption{OpenMC comparison with MCNP5 using ENDF/B-VII.0.}
\label{tab:library_mcnp}
\footnotesize
\begin{tabular}{@{}lccc@{}}
\toprule
Name & OpenMC $\alpha_{\mathrm{dc}}$ & MCNP5 $\alpha_{\mathrm{dc}}$~\cite{mosteller2011} & Units (s$^{-1}$) \\
\midrule
Godiva       & $-1.1519 \pm 0.0042$ & $-1.130 \pm 0.010$ & $\times 10^{6}$ \\
Flattop-25   & $-0.3944 \pm 0.0014$ & $-0.397 \pm 0.002$ & $\times 10^{6}$ \\
Zeus-1       & $-0.3615 \pm 0.0013$ & $-0.363 \pm 0.002$ & $\times 10^{4}$ \\
Zeus-5       & $-10.7043 \pm 0.0399$ & $-10.760 \pm 0.070$ & $\times 10^{4}$ \\
Zeus-6       & $-4.1009 \pm 0.0148$ & $-4.140 \pm 0.030$ & $\times 10^{4}$ \\
Big Ten      & $-0.1182 \pm 0.0004$ & $-0.118 \pm 0.001$ & $\times 10^{6}$ \\
Jezebel-Pu   & $-0.6437 \pm 0.0041$ & $-0.650 \pm 0.010$ & $\times 10^{6}$ \\
Flattop-Pu   & $-0.2093 \pm 0.0011$ & $-0.210 \pm 0.003$ & $\times 10^{6}$ \\
Thor         & $-0.2091 \pm 0.0013$ & $-0.200 \pm 0.010$ & $\times 10^{6}$ \\
Jezebel-233  & $-1.0789 \pm 0.0054$ & $-1.080 \pm 0.010$ & $\times 10^{6}$ \\
Flattop-23   & $-0.2950 \pm 0.0014$ & $-0.302 \pm 0.004$ & $\times 10^{6}$ \\
\bottomrule
\end{tabular}
\end{table}

\begin{table}[htbp]
\centering
\caption{OpenMC comparison with OpenMC(TD) using ENDF/B-VII.1.}
\label{tab:library_td}
\footnotesize
\begin{tabular}{@{}lccc@{}}
\toprule
Name & OpenMC $\alpha_{\mathrm{dc}}$ & OpenMC(TD) $\alpha_{\mathrm{dc}}$~\cite{romero2022} & Units (s$^{-1}$) \\
\midrule
Godiva       & $-1.1480 \pm 0.0041$ & $-1.110 \pm 0.020$ & $\times 10^{6}$ \\
Big Ten      & $-0.1179 \pm 0.0004$ & $-0.110 \pm 0.002$ & $\times 10^{6}$ \\
Jezebel-Pu   & $-0.6425 \pm 0.0040$ & $-0.628 \pm 0.028$ & $\times 10^{6}$ \\
Flattop-23   & $-0.2932 \pm 0.0014$ & $-0.268 \pm 0.016$ & $\times 10^{6}$ \\
RA-6         & $-1.7203 \pm 0.0050$ & $-1.724 \pm 0.027$ & $\times 10^{2}$ \\
STACY-029    & $-1.2179 \pm 0.0036$ & $-1.154 \pm 0.014$ & $\times 10^{2}$ \\
STACY-033    & $-1.1569 \pm 0.0033$ & $-1.101 \pm 0.012$ & $\times 10^{2}$ \\
STACY-046    & $-1.0575 \pm 0.0031$ & $-1.005 \pm 0.015$ & $\times 10^{2}$ \\
STACY-030    & $-1.2532 \pm 0.0037$ & $-1.197 \pm 0.014$ & $\times 10^{2}$ \\
STACY-125    & $-1.5214 \pm 0.0044$ & $-1.464 \pm 0.017$ & $\times 10^{2}$ \\
\bottomrule
\end{tabular}
\end{table}

\subsection{Note on the Zeus Benchmarks}\label{sec:zeus}

The three Zeus benchmarks (Zeus-1, Zeus-5, and Zeus-6) are consistent outliers in the Rossi-$\alpha$ validation suite~\cite{mosteller2011}. All three are copper-reflected HEU systems that were assembled on the Comet vertical lift machine at the Los Alamos Critical Experiments Facility. Zeus-5 shows the largest discrepancy: the measured $\alpha_{\mathrm{dc}}$ is $(-7.960 \pm 0.080) \times 10^4$~s$^{-1}$, while the OpenMC value from this work is $(-10.520 \pm 0.038) \times 10^4$~s$^{-1}$, consistent with the MCNP5 value of $(-10.760 \pm 0.070) \times 10^4$~s$^{-1}$ reported by Mosteller and Kiedrowski~\cite{mosteller2011}. Zeus-1 and Zeus-6 show smaller but qualitatively similar overpredictions. The source of this discrepancy is not well understood.

An updated Rossi-$\alpha$ measurement of Zeus-5 was published by Sanchez et al.~\cite{sanchez2015}. The unmoderated Zeus core was reassembled on the Comet machine at the National Criticality Experiments Research Center with a plate loading and in-core hardware that differed slightly from the HMF-073 benchmark configuration. The delayed-critical alpha eigenvalue inferred from measurements at different separation distances between core halves was $(-8.99 \pm 0.08) \times 10^4$~s$^{-1}$, compared with the historical value of $(-7.96 \pm 0.08) \times 10^4$~s$^{-1}$ in Table~\ref{tab:alpha_comparison}. The updated measurement improves the C/E for the OpenMC $\alpha_{\mathrm{dc}}$ from 1.32 to 1.17, consistent with the approximately 11\% overprediction reported for MCNP5 by McKenzie~\cite{mckenzie2019}.

The recent CERBERUS experiment campaign~\cite{kostelac2026} provides further context. CERBERUS reused the HEU fuel plates and copper reflector from the original Zeus experiments in a new configuration on the Comet machine. Kostelac et al.\ measured the prompt neutron decay constant via the cross power spectral density method using in-core $^3$He proportional counters and ex-core organic scintillators at delayed critical and four subcritical states. The extrapolated $\alpha_{\mathrm{dc}}$ values were $(-5.386 \pm 0.063) \times 10^4$~s$^{-1}$ ($^3$He) and $(-5.188 \pm 0.307) \times 10^4$~s$^{-1}$ (scintillators), while an MCNP6.2 calculation using ENDF/B-VIII.0 yielded $(-6.254 \pm 0.018) \times 10^4$~s$^{-1}$, an overprediction of 16\% and 21\%, respectively. During the 3.5-hour delayed-critical measurement, operator intervention was required multiple times to stabilize the diverging neutron population by adjusting the platen position in increments of approximately 1~mil (0.0254~mm)~\cite{kostelac2026}, illustrating the difficulty of holding these assemblies at delayed critical.

The benchmark configuration of Zeus-5 was slightly supercritical, with a documented experimental $k_{\mathrm{eff}}$ of $1.0008 \pm 0.0015$~\cite{icsbep}, consistent with the calculated value of $1.00204 \pm 0.00005$. The most direct comparison with the measured Rossi-$\alpha$ is obtained by bringing the model to delayed critical. A series of calculations was performed in which the gap between the upper and lower halves of the assembly was opened in 0.25~mm increments from the closed benchmark configuration. The results are presented in Table~\ref{tab:zeus5_gap}. As the gap opens, the system becomes less reactive and $\alpha_{\mathrm{static}}$ decreases, passing through $\alpha_{\mathrm{dc}}$ where $k_{\mathrm{eff}} = 1$, while $\alpha_{\mathrm{dc}}$ itself remains at approximately $-10.5 \times 10^4$~s$^{-1}$. The overprediction is therefore not attributable to the reactivity bias of the model.
\begin{table}[htbp]
\centering
\footnotesize
\begin{threeparttable}
\caption{Calculated Zeus-5 gap sensitivity.$^{**}$}
\label{tab:zeus5_gap}
\begin{tabular}{@{}lccccc@{}}
\toprule
Gap & OpenMC $k_{\mathrm{eff}}^*$ & OpenMC $k_p^*$ & OpenMC $\alpha_{\mathrm{dc}}$ & OpenMC $\alpha_{\mathrm{static}}$ & Units (s$^{-1}$) \\
\midrule
Benchmark & 1.00204 & 0.99526 & $-10.520 \pm 0.038$ & $-7.370 \pm 0.086$ & $\times 10^4$ \\
0.25 mm & 1.00148 & 0.99474 & $-10.450 \pm 0.038$ & $-8.162 \pm 0.084$ & $\times 10^4$ \\
0.5 mm & 1.00070 & 0.99395 & $-10.456 \pm 0.038$ & $-9.371 \pm 0.086$ & $\times 10^4$ \\
\rowcolor{orange!15}
0.75 mm & 1.00007 & 0.99326 & $-10.533 \pm 0.039$ & $-10.421 \pm 0.086$ & $\times 10^4$ \\
1.0 mm & 0.99940 & 0.99265 & $-10.409 \pm 0.038$ & $-11.328 \pm 0.085$ & $\times 10^4$ \\
1.25 mm & 0.99858 & 0.99186 & $-10.360 \pm 0.038$ & $-12.525 \pm 0.085$ & $\times 10^4$ \\
1.5 mm & 0.99788 & 0.99111 & $-10.432 \pm 0.038$ & $-13.664 \pm 0.087$ & $\times 10^4$ \\
\bottomrule
\end{tabular}
\begin{tablenotes}[flushleft]
\footnotesize
\item[]$^*$Statistical uncertainty on $k_{\mathrm{eff}}$ is $< 0.0001$ for all cases.
\item[]$^{**}$The highlighted (orange) row is the point nearest delayed critical, where $\alpha_{\mathrm{static}}$ is closest to $\alpha_{\mathrm{dc}}$ ($k_{\mathrm{eff}} = 1$).
\end{tablenotes}
\end{threeparttable}
\end{table}

\subsection{SHE-8 Subcritical Extrapolation}

\begin{figure}[htbp]
\centering
\includegraphics[width=0.78\textwidth]{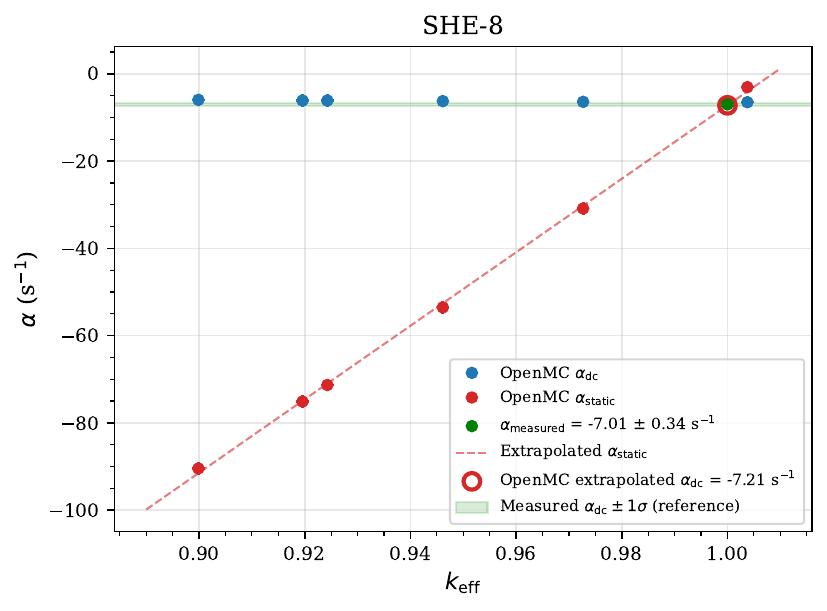}
\caption{SHE-8: OpenMC $\alpha_{\mathrm{dc}}$ and $\alpha_{\mathrm{static}}$ as a function of $k_{\mathrm{eff}}$. The measured Rossi-$\alpha$ is a single delayed-critical quantity, drawn at $k_{\mathrm{eff}} = 1$; the shaded green band extends it across the panel for visual reference only. The dashed line is the linear extrapolation of the calculated $\alpha_{\mathrm{static}}$ to $k_{\mathrm{eff}} = 1$.}
\label{fig:she8}
\end{figure}

SHE-8 is a 240-cm-tall hexagonal graphite core loaded with LEU fuel rods and graphite rods, with a prompt neutron generation time on the order of $10^{-3}$~s. The subcritical configurations originate from control rod worth measurements performed in support of the Japanese Experimental Very High Temperature Reactor: gray and black absorber rods were inserted into the core, and the rod worth was obtained from the change in the prompt neutron decay constant measured by Takano et al.~\cite{takano1985} using the pulsed neutron source (PNS) method. Modeling these configurations, where $k_{\mathrm{eff}}$ is reduced from approximately 1.004 to 0.900, allows the behavior of $\alpha_{\mathrm{dc}}$ and $\alpha_{\mathrm{static}}$ to be studied as a function of subcriticality. The results are presented in Table~\ref{tab:she8_subcrit} and Fig.~\ref{fig:she8}.

At the critical configuration, the calculated $\alpha_{\mathrm{dc}}$ of $-6.501$~s$^{-1}$ underpredicts the measured value of $-7.01 \pm 0.34$~s$^{-1}$~\cite{takano1985} by 7.3\%. As $k_{\mathrm{eff}}$ decreases from approximately 1.004 to 0.900, $\alpha_{\mathrm{static}}$ decreases from $-3.05$ to $-90.44$~s$^{-1}$, while $\alpha_{\mathrm{dc}}$ increases monotonically from $-6.501$ to $-5.945$~s$^{-1}$ (Table~\ref{tab:she8_subcrit}), a change of 8.6\% over a 10\% range in $k_{\mathrm{eff}}$. A linear extrapolation of $\alpha_{\mathrm{static}}$ to $k_{\mathrm{eff}} = 1.0$ gives $-7.21$~s$^{-1}$, which is within 2.8\% of the measured value and in better agreement than the $\alpha_{\mathrm{dc}}$ calculated directly at the critical configuration.

\begin{table}[htbp]
\centering
\caption{SHE-8 subcritical series. Extrapolated $\alpha_{\mathrm{dc}} = -7.21$~s$^{-1}$.}
\label{tab:she8_subcrit}
\small
\begin{tabular}{@{}lccccc@{}}
\toprule
Configuration & OpenMC $k_{\mathrm{eff}}^*$ & $\alpha_{\mathrm{measured}}$~\cite{takano1985} & OpenMC $\alpha_{\mathrm{static}}$ & OpenMC $\alpha_{\mathrm{dc}}$ & Units (s$^{-1}$) \\
\midrule
Critical & 1.00377 & $-7.01 \pm 0.34$ & $-3.05 \pm 0.06$ & $-6.501 \pm 0.019$ & $\times 10^0$ \\
Gray absorber ($\times$1) & 0.97269 & $-35.92 \pm 0.29$ & $-30.86 \pm 0.06$ & $-6.401 \pm 0.019$ & $\times 10^0$ \\
Gray absorber ($\times$3) & 0.94614 & $-55.62 \pm 0.37$ & $-53.53 \pm 0.07$ & $-6.252 \pm 0.019$ & $\times 10^0$ \\
Gray absorber ($\times$5) & 0.92428 & $-70.62 \pm 0.70$ & $-71.32 \pm 0.07$ & $-6.110 \pm 0.019$ & $\times 10^0$ \\
Gray absorber ($\times$7) & 0.89988 & $-84.42 \pm 0.78$ & $-90.44 \pm 0.08$ & $-5.945 \pm 0.019$ & $\times 10^0$ \\
Black absorber & 0.91959 & $-78.70 \pm 0.92$ & $-75.11 \pm 0.07$ & $-6.107 \pm 0.019$ & $\times 10^0$ \\
\bottomrule
\multicolumn{6}{@{}l}{\footnotesize $^*$Statistical uncertainty on $k_{\mathrm{eff}}$ is $< 0.0001$ for all cases.} \\
\end{tabular}
\end{table}

\subsection{STACY Subcritical Extrapolation}

\begin{figure}[htbp]
\centering
\includegraphics[width=\textwidth]{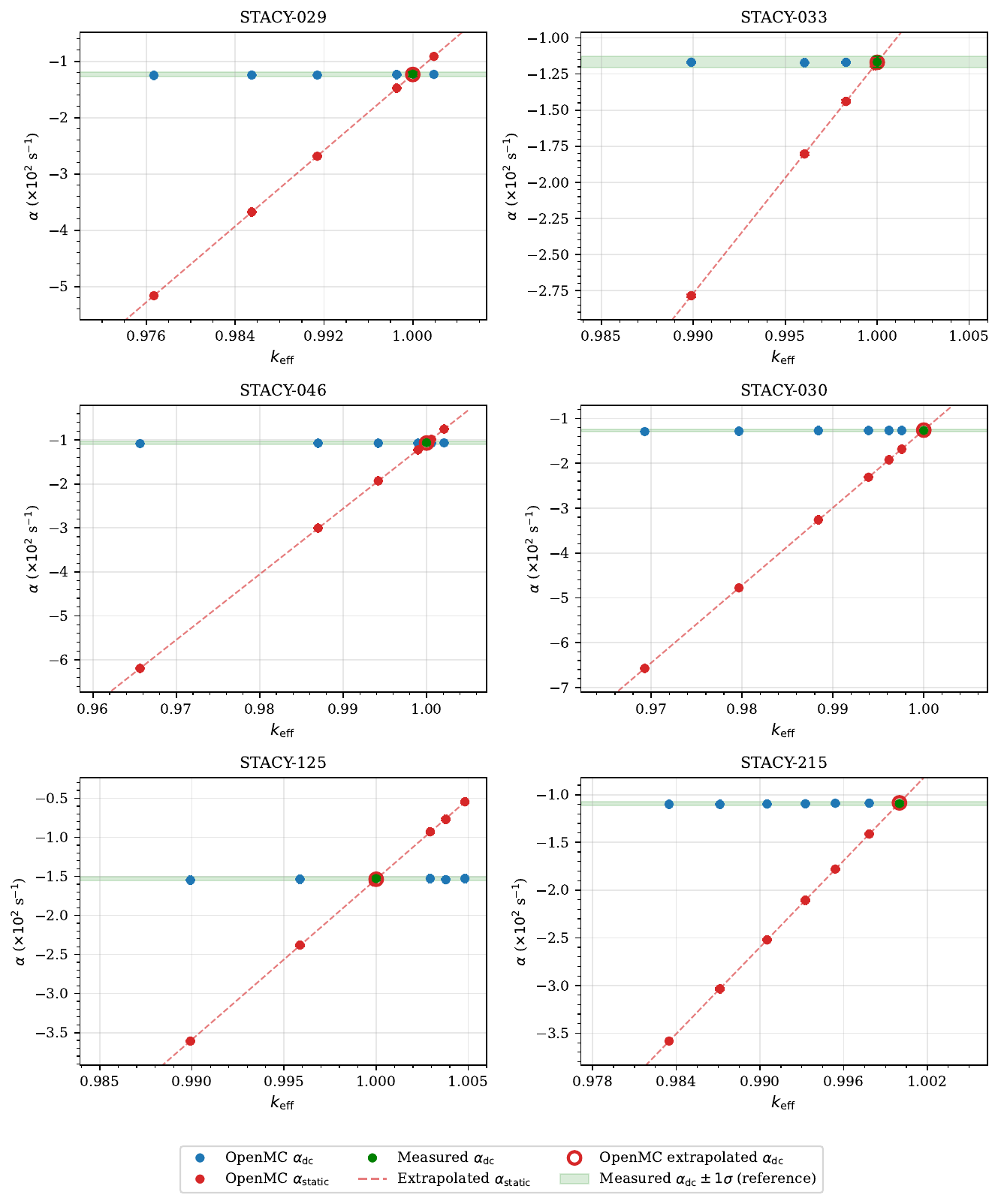}
\caption{STACY subcritical extrapolation. In each panel the measured $\alpha_{\mathrm{dc}}$~\cite{tonoike2002} is a single value, drawn at $k_{\mathrm{eff}} = 1$ and extended across the panel by the shaded green band for visual reference only. The dashed lines are linear extrapolations of the calculated $\alpha_{\mathrm{static}}$ to $k_{\mathrm{eff}} = 1$.}
\label{fig:stacy}
\end{figure}

Subcritical cases for each of the six STACY configurations were generated by progressively reducing the solution height. Tonoike et al.~\cite{tonoike2002} measured the prompt neutron decay constant at various subcritical solution levels using the PNS method. Where an OpenMC subcritical configuration matches a measured solution level, the measured prompt decay constants are included in Tables~\ref{tab:stacy029} through~\ref{tab:stacy215}. The measured value listed for each parent configuration is $\alpha_{\mathrm{dc}}$, obtained by extrapolating the subcritical measurements to zero reactivity. The extrapolated $\alpha_{\mathrm{dc}}$ in each table caption is the OpenMC equivalent: a linear fit of the calculated $\alpha_{\mathrm{static}}$ values, evaluated at $k_{\mathrm{eff}} = 1$. Figure~\ref{fig:stacy} shows the subcritical extrapolation for all six STACY configurations.

For each series, OpenMC $\alpha_{\mathrm{dc}}$ remains approximately constant, varying by no more than 2\%. The OpenMC $\alpha_{\mathrm{static}}$ values decrease with subcriticality and follow the trend of the PNS measurements of Tonoike et al.~\cite{tonoike2002}. When the system is close to delayed critical, $\alpha_{\mathrm{static}}$ and $\alpha_{\mathrm{measured}}$ are offset by the reactivity bias of the model; the agreement between $\alpha_{\mathrm{static}}$ and $\alpha_{\mathrm{measured}}$ improves as the system is driven subcritical. For STACY-029, where the calculated $k_{\mathrm{eff}}$ is 1.00190, the difference falls from 26\% at the parent configuration to 4\% at the most subcritical case.

\begin{table}[htbp]
\centering\caption{STACY-029 subcritical series. Extrapolated $\alpha_{\mathrm{dc}} = -1.234 \times 10^2$~s$^{-1}$.}\label{tab:stacy029}\footnotesize
\begin{tabular}{@{}lrccccc@{}}
\toprule
Case & Level (mm) & OpenMC $k_{\mathrm{eff}}^*$ & $\alpha_{\mathrm{measured}}$~\cite{tonoike2002} & OpenMC $\alpha_{\mathrm{static}}$ & OpenMC $\alpha_{\mathrm{dc}}$ & Units (s$^{-1}$) \\
\midrule
LST004-2 & 467.0 & 1.00190 & $-1.227 \pm 0.041$ & $-0.913 \pm 0.010$ & $-1.231 \pm 0.003$ & $\times 10^2$ \\
LST004-2A & 462.0 & 1.00022 & $-1.480 \pm 0.053$ & $-1.197 \pm 0.010$ & $-1.233 \pm 0.003$ & $\times 10^2$ \\
LST004-2B & 456.5 & 0.99856 & $-1.805 \pm 0.046$ & $-1.476 \pm 0.010$ & $-1.234 \pm 0.004$ & $\times 10^2$ \\
LST004-2C & 437.0 & 0.99140 & $-2.959 \pm 0.045$ & $-2.685 \pm 0.010$ & $-1.244 \pm 0.004$ & $\times 10^2$ \\
LST004-2D & 422.3 & 0.98549 & $-3.923 \pm 0.072$ & $-3.676 \pm 0.010$ & $-1.244 \pm 0.004$ & $\times 10^2$ \\
LST004-2E & 402.0 & 0.97667 & $-5.385 \pm 0.155$ & $-5.162 \pm 0.011$ & $-1.250 \pm 0.004$ & $\times 10^2$ \\
\bottomrule
\multicolumn{7}{@{}l}{\footnotesize $^*$Statistical uncertainty on $k_{\mathrm{eff}}$ is $< 0.0001$ for all cases.} \\
\end{tabular}
\end{table}

\begin{table}[htbp]
\centering\caption{STACY-033 subcritical series. Extrapolated $\alpha_{\mathrm{dc}} = -1.169 \times 10^2$~s$^{-1}$.}\label{tab:stacy033}\footnotesize
\begin{tabular}{@{}lrccccc@{}}
\toprule
Case & Level (mm) & OpenMC $k_{\mathrm{eff}}^*$ & $\alpha_{\mathrm{measured}}$~\cite{tonoike2002} & OpenMC $\alpha_{\mathrm{static}}$ & OpenMC $\alpha_{\mathrm{dc}}$ & Units (s$^{-1}$) \\
\midrule
LST004-3 & 529.3 & 0.99985 & $-1.167 \pm 0.039$ & $-1.191 \pm 0.010$ & $-1.168 \pm 0.003$ & $\times 10^2$ \\
LST004-3A & 523.1 & 0.99831 & $-1.411 \pm 0.040$ & $-1.440 \pm 0.010$ & $-1.169 \pm 0.003$ & $\times 10^2$ \\
LST004-3B & 514.2 & 0.99604 & $-1.764 \pm 0.030$ & $-1.804 \pm 0.010$ & $-1.172 \pm 0.003$ & $\times 10^2$ \\
LST004-3C & 491.1 & 0.98989 & $-2.779 \pm 0.055$ & $-2.785 \pm 0.010$ & $-1.168 \pm 0.003$ & $\times 10^2$ \\
\bottomrule
\multicolumn{7}{@{}l}{\footnotesize $^*$Statistical uncertainty on $k_{\mathrm{eff}}$ is $< 0.0001$ for all cases.} \\
\end{tabular}
\end{table}

\begin{table}[htbp]
\centering\caption{STACY-046 subcritical series. Extrapolated $\alpha_{\mathrm{dc}} = -1.070 \times 10^2$~s$^{-1}$.}\label{tab:stacy046}\footnotesize
\begin{tabular}{@{}lrccccc@{}}
\toprule
Case & Level (mm) & OpenMC $k_{\mathrm{eff}}^*$ & $\alpha_{\mathrm{measured}}$~\cite{tonoike2002} & OpenMC $\alpha_{\mathrm{static}}$ & OpenMC $\alpha_{\mathrm{dc}}$ & Units (s$^{-1}$) \\
\midrule
LST004-5 & 785.6 & 1.00210 & $-1.062 \pm 0.037$ & $-0.753 \pm 0.009$ & $-1.065 \pm 0.003$ & $\times 10^2$ \\
LST004-5A & 770.0 & 1.00058 & $-1.278 \pm 0.094$ & $-0.986 \pm 0.009$ & $-1.072 \pm 0.003$ & $\times 10^2$ \\
LST004-5B & 750.1 & 0.99898 & $-1.532 \pm 0.028$ & $-1.225 \pm 0.009$ & $-1.074 \pm 0.003$ & $\times 10^2$ \\
LST004-5C & 704.8 & 0.99422 & $-2.251 \pm 0.040$ & $-1.931 \pm 0.009$ & $-1.072 \pm 0.003$ & $\times 10^2$ \\
LST004-5D & 648.8 & 0.98700 & $-3.312 \pm 0.096$ & $-3.005 \pm 0.009$ & $-1.073 \pm 0.003$ & $\times 10^2$ \\
LST004-5E & 530.4 & 0.96565 & $-6.458 \pm 0.336$ & $-6.195 \pm 0.009$ & $-1.082 \pm 0.003$ & $\times 10^2$ \\
\bottomrule
\multicolumn{7}{@{}l}{\footnotesize $^*$Statistical uncertainty on $k_{\mathrm{eff}}$ is $< 0.0001$ for all cases.} \\
\end{tabular}
\end{table}

\begin{table}[htbp]
\centering\caption{STACY-030 subcritical series. Extrapolated $\alpha_{\mathrm{dc}} = -1.261 \times 10^2$~s$^{-1}$.}\label{tab:stacy030}\footnotesize
\begin{tabular}{@{}lrccccc@{}}
\toprule
Case & Level (mm) & OpenMC $k_{\mathrm{eff}}^*$ & $\alpha_{\mathrm{measured}}$~\cite{tonoike2002} & OpenMC $\alpha_{\mathrm{static}}$ & OpenMC $\alpha_{\mathrm{dc}}$ & Units (s$^{-1}$) \\
\midrule
LST007-2 & 542.0 & 0.99758 & $-1.268 \pm 0.029$ & $-1.685 \pm 0.011$ & $-1.270 \pm 0.004$ & $\times 10^2$ \\
LST007-2A & 536.3 & 0.99617 & $-1.495 \pm 0.032$ & $-1.923 \pm 0.011$ & $-1.268 \pm 0.004$ & $\times 10^2$ \\
LST007-2B & 527.2 & 0.99392 & $-1.867 \pm 0.042$ & $-2.311 \pm 0.011$ & $-1.270 \pm 0.004$ & $\times 10^2$ \\
LST007-2C & 506.4 & 0.98839 & $-2.772 \pm 0.045$ & $-3.262 \pm 0.011$ & $-1.273 \pm 0.004$ & $\times 10^2$ \\
LST007-2D & 477.3 & 0.97965 & $-4.247 \pm 0.055$ & $-4.778 \pm 0.011$ & $-1.284 \pm 0.004$ & $\times 10^2$ \\
LST007-2E & 447.2 & 0.96928 & $-6.057 \pm 0.182$ & $-6.572 \pm 0.011$ & $-1.293 \pm 0.004$ & $\times 10^2$ \\
\bottomrule
\multicolumn{7}{@{}l}{\footnotesize $^*$Statistical uncertainty on $k_{\mathrm{eff}}$ is $< 0.0001$ for all cases.} \\
\end{tabular}
\end{table}

\begin{table}[htbp]
\centering\caption{STACY-125 subcritical series. Extrapolated $\alpha_{\mathrm{dc}} = -1.536 \times 10^2$~s$^{-1}$.}\label{tab:stacy125}\footnotesize
\begin{tabular}{@{}lrccccc@{}}
\toprule
Case & Level (mm) & OpenMC $k_{\mathrm{eff}}^*$ & $\alpha_{\mathrm{measured}}$~\cite{tonoike2002} & OpenMC $\alpha_{\mathrm{static}}$ & OpenMC $\alpha_{\mathrm{dc}}$ & Units (s$^{-1}$) \\
\midrule
LST016-3 & 513.7 & 1.00482 & $-1.528 \pm 0.026$ & $-0.545 \pm 0.013$ & $-1.528 \pm 0.004$ & $\times 10^2$ \\
LST016-3A & 509.9 & 1.00378 & $-1.704 \pm 0.029$ & $-0.771 \pm 0.013$ & $-1.542 \pm 0.004$ & $\times 10^2$ \\
LST016-3B & 506.0 & 1.00294 & $-1.925 \pm 0.048$ & $-0.929 \pm 0.013$ & $-1.528 \pm 0.004$ & $\times 10^2$ \\
LST016-3C & 494.6 & 0.99985 & $-2.522 \pm 0.029$ & $-1.567 \pm 0.013$ & $-1.535 \pm 0.004$ & $\times 10^2$ \\
LST016-3D & 480.3 & 0.99586 & $-3.324 \pm 0.030$ & $-2.380 \pm 0.013$ & $-1.536 \pm 0.004$ & $\times 10^2$ \\
LST016-3E & 460.4 & 0.98992 & $-4.522 \pm 0.066$ & $-3.608 \pm 0.013$ & $-1.549 \pm 0.004$ & $\times 10^2$ \\
\bottomrule
\multicolumn{7}{@{}l}{\footnotesize $^*$Statistical uncertainty on $k_{\mathrm{eff}}$ is $< 0.0001$ for all cases.} \\
\end{tabular}
\end{table}

\begin{table}[htbp]
\centering\caption{STACY-215 subcritical series. Extrapolated $\alpha_{\mathrm{dc}} = -1.085 \times 10^2$~s$^{-1}$.}\label{tab:stacy215}\footnotesize
\begin{tabular}{@{}lrccccc@{}}
\toprule
Case & Level (mm) & OpenMC $k_{\mathrm{eff}}^*$ & $\alpha_{\mathrm{measured}}$~\cite{tonoike2002} & OpenMC $\alpha_{\mathrm{static}}$ & OpenMC $\alpha_{\mathrm{dc}}$ & Units (s$^{-1}$) \\
\midrule
LST021-1 & 439.8 & 0.99784 & $-1.092 \pm 0.018$ & $-1.412 \pm 0.009$ & $-1.088 \pm 0.003$ & $\times 10^2$ \\
LST021-1A & 433.6 & 0.99540 & $-1.450 \pm 0.018$ & $-1.779 \pm 0.009$ & $-1.089 \pm 0.003$ & $\times 10^2$ \\
LST021-1B & 428.6 & 0.99325 & $-1.764 \pm 0.019$ & $-2.106 \pm 0.009$ & $-1.094 \pm 0.003$ & $\times 10^2$ \\
LST021-1C & 421.9 & 0.99051 & $-2.189 \pm 0.021$ & $-2.522 \pm 0.009$ & $-1.097 \pm 0.003$ & $\times 10^2$ \\
LST021-1D & 414.4 & 0.98713 & $-2.668 \pm 0.019$ & $-3.035 \pm 0.009$ & $-1.101 \pm 0.003$ & $\times 10^2$ \\
LST021-1E & 406.5 & 0.98348 & $-3.211 \pm 0.023$ & $-3.583 \pm 0.009$ & $-1.100 \pm 0.003$ & $\times 10^2$ \\
\bottomrule
\multicolumn{7}{@{}l}{\footnotesize $^*$Statistical uncertainty on $k_{\mathrm{eff}}$ is $< 0.0001$ for all cases.} \\
\end{tabular}
\end{table}

\section{Conclusions}\label{sec:conclusions}

Alpha eigenvalue calculations were performed for 21 delayed-critical benchmark experiments and 33 subcritical configurations using a modified version of OpenMC. The effective delayed neutron fraction and prompt neutron lifetime were calculated using the iterated fission probability method. Two forms of the alpha eigenvalue were derived from these quantities using the point kinetics approximation: the delayed-critical alpha eigenvalue ($\alpha_{\mathrm{dc}} = -\beta_{\mathrm{eff}} / \ell_p$) and the static alpha eigenvalue ($\alpha_{\mathrm{static}} = (k_p - 1) / \ell_p$), which reflects the actual reactivity state of the model.

The calculated $\alpha_{\mathrm{dc}}$ values agree with the measured Rossi-$\alpha$ values to within 0.4\% for the six STACY thermal solution benchmarks, with C/E ratios ranging from 0.996 to 1.004. For fast metal systems, agreement is generally within 5--10\%, comparable to previously published results from MCNP5 and Serpent~2, although these systems are consistently overpredicted; the largest differences are for the fertile-reflected Flattop-23 and Thor (13\% and 11\%). The three Zeus copper-reflected benchmarks remain outliers. The updated measurement of Sanchez et al.~\cite{sanchez2015} improves the Zeus-5 C/E for $\alpha_{\mathrm{dc}}$ from 1.32 to 1.17, and a gap sensitivity study shows that the overprediction is not attributable to the reactivity bias of the model.

Subcritical extrapolation studies for SHE-8 and the STACY series show that the IFP-based kinetics parameters track the neutronic characteristics of the system rather than its reactivity state. In both cases, $\alpha_{\mathrm{dc}}$ remains approximately constant as the system is driven subcritical, while $\alpha_{\mathrm{static}}$ decreases, as expected from the point kinetics approximation. For the STACY series, the calculated $\alpha_{\mathrm{static}}$ values follow the trend of the pulsed neutron source measurements of Tonoike et al.~\cite{tonoike2002}, and the agreement between calculation and measurement improves with subcriticality.

Comparison between OpenMC and previously published results from MCNP5, Serpent~2, and OpenMC(TD) shows broad consistency across all four codes. Supplementary calculations with ENDF/B-VII.0 and ENDF/B-VII.1 cross sections confirm that the nuclear data library has only a small effect on $\alpha_{\mathrm{dc}}$; the remaining differences are attributable mainly to the methods used to calculate the underlying kinetics parameters, particularly the fitted $\Lambda_p$ used by OpenMC(TD).

\section*{Acknowledgments}

This work was performed under the auspices of the U.S.\ Department of Energy by Lawrence Livermore National Laboratory under Contract DE-AC52-07NA27344. The corresponding author would like to thank Greg Spriggs for generously sharing his time and insight on Rossi-$\alpha$ and $\beta_{\mathrm{eff}}$ measurements of critical experiments. His work has long been an inspiration for this line of research.

\end{document}